\documentclass[slac_one]{revtex4}
\usepackage{graphicx}
\usepackage{fancyhdr}
\newcommand{\pt}{\ensuremath{p_{T}}}
\newcommand{\et}{\ensuremath{E_{T}}}
\newcommand{\met}{\ensuremath{E_{T}\!\!\!\!\!\!/}}
\newcommand{\gev}{\ensuremath{\rm GeV}}

\newcommand{\fbarn} {\ensuremath{\mathrm{fb^{-1}}}}
\newcommand{\g} {\ensuremath{\gamma}}
\def \kaione {{\tilde \chi}_1^0}
\def \kaioneplus {{\tilde \chi}_1^+}
\def \kaitwo {{\tilde \chi}_2^0}
\def\stop{\tilde{t}}

\begin{document}

\title{Search for Anomalous Production of Photon, $b$-jet, and Missing 
Transverse Energy at CDF} 

%

\author{Shin-Shan Yu (for the CDF Collaboration)}
\affiliation{Fermi National Accelerator Laboratory, P.O.~Box~500, Batavia, IL~60510, USA}
\begin{abstract}
 We report the results on two signature-based searches for new physics 
 using 1.9--2.0~\fbarn\ of data collected at the CDF experiment. 
 Both analyses look in events containing a photon, a $b$-tagged jet, and 
 missing transverse energy. The first search requires an additional jet. The 
 second search requires an extra electron or muon. 
 No significant excess of events over the Standard Model 
 prediction is observed. We also describe the ``CES/CPR'' method which 
 is used to estimate the amount of mis-identified photons. 
\end{abstract}

\maketitle

\thispagestyle{fancy}


\section{\boldmath SEARCH FOR ANOMALOUS PRODUCTION OF $\g bj\met$ 
            \label{sec:gbjm}}
\begin{figure}
\begin{tabular}{cc}
\includegraphics[width=0.4\textwidth]{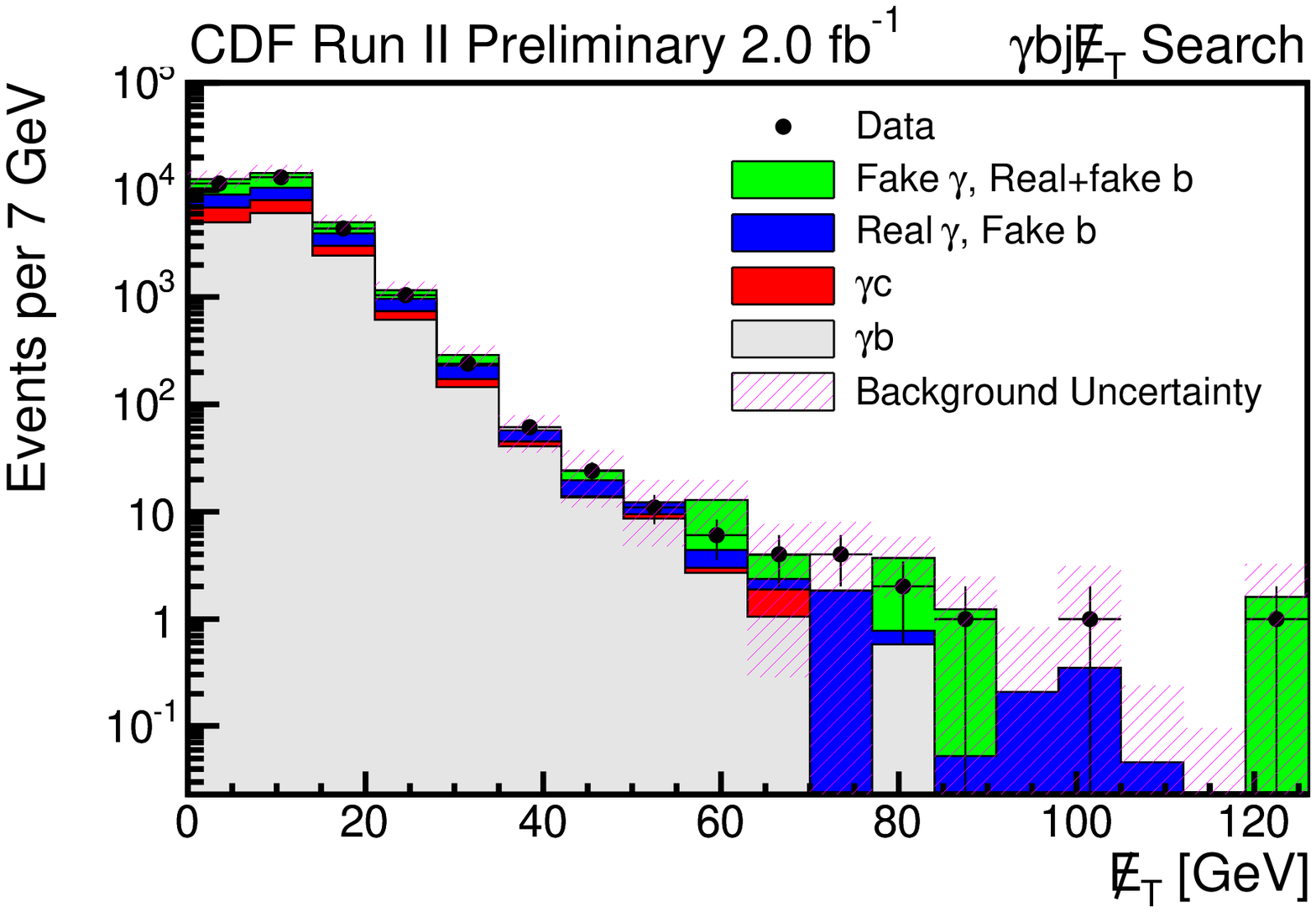} &   
\includegraphics[width=0.4\textwidth]{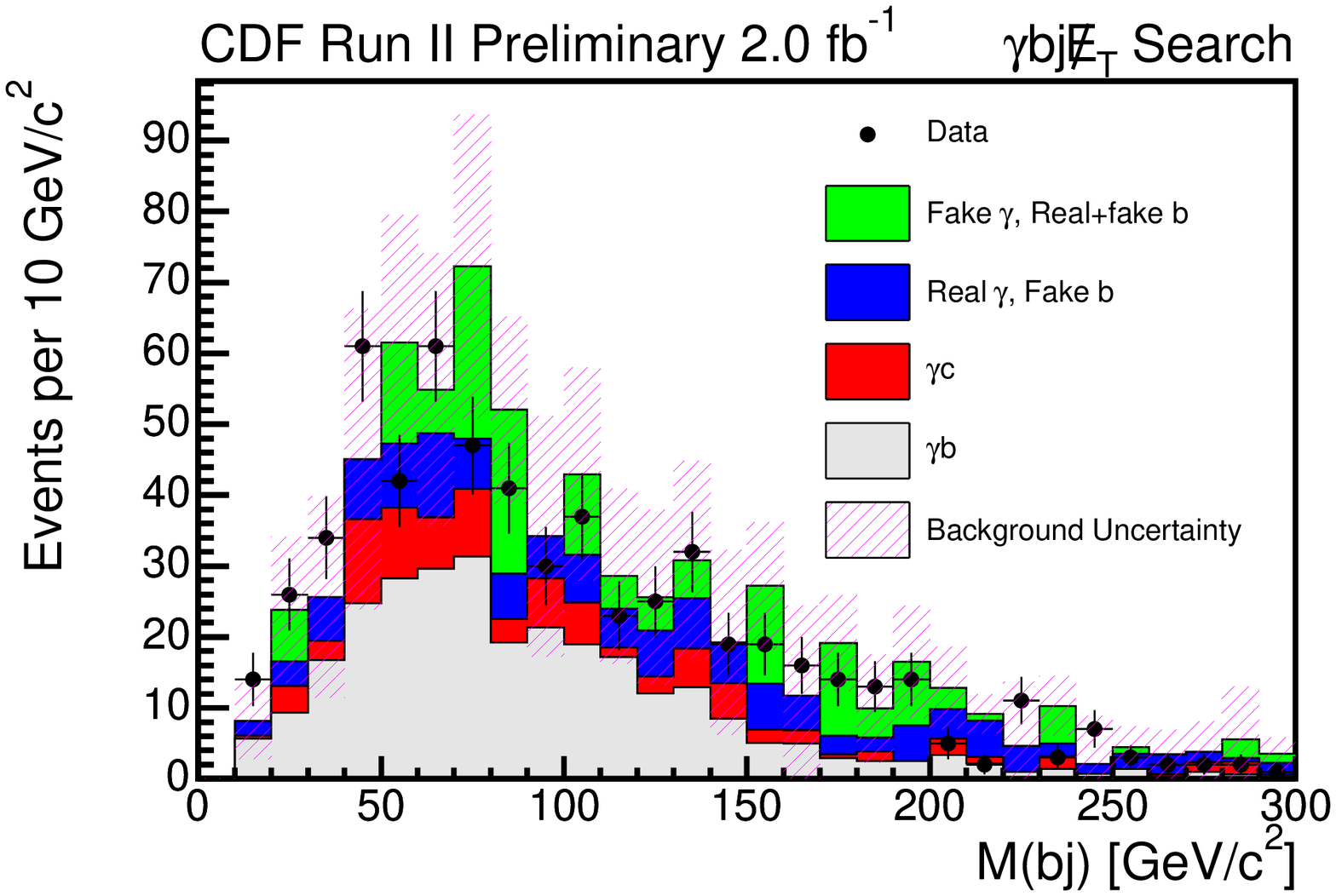} \\
\end{tabular}  
	\caption{\label{fig:gbjmet}
	CDF search for anomalous production of $\g bj\met\;$:
	the $\met\;$ (left) and $M_{bj}$ (right) distributions observed 
	(markers) and background prediction (filled histograms). 
	The hatched-region indicates the total 
	uncertainty on the predicted background in each bin.	
	}
\end{figure}

\unboldmath
The CDF collaboration has performed a signature-based search in the inclusive 
$\g bj\met\;$ final state using 2.0~\fbarn\ of data. The $\g bj\met\;$ 
signature raised great interest for two main reasons. First, this final state 
has been predicted by several SUSY models\footnote{These models had 
been proposed to explain the CDF $ee\g\g\met\;$ event observed in Run I~\cite{Toback_all}.}\cite{Kane:1996ny,Kane}, {\it e.g.}, the production of a chargino and a 
neutralino, when $\kaitwo$ is photino-like and the LSP $\kaione$ is 
Higgsino-like, via the decay chain: 
$\kaioneplus \kaitwo \rightarrow  (\bar{b}\stop )(\gamma\kaione) \rightarrow 
(\bar{b}c \kaione)(\gamma\kaione)  \rightarrow (\gamma \bar{b} c \met\;)
$. Second, the dominant backgrounds are mis-identifications of either the 
photon or the $b$-quark candidates and mismeasurements of the jet energy which 
induce $\met\;$ not associated with unobserved neutral particles (fake $\met\;$).  
The SM processes which produce real $\g bj\met\;$ are expected to contribute at
 most 2\%. Therefore, a significant excess in data will be an indication of new
 physics. Events are required to have 
a central\footnote{Throughout this document, all central objects have 
detector pseudo-rapidity $\left|\eta^\mathrm{det}\right| < 1.1$.} photon with 
transverse energy $\et>25~\gev$, at least two jets with $\et>15~\gev$ and 
$\left|\eta^\mathrm{det}\right| < 2.0$, at least one of the jets must be 
identified as originating from a $b$ quark (``$b$-tagged'') using the tight 
SECVTX algorithm~\cite{SECVTX}, and missing transverse energy $\met\; > 25~\gev$.
 Figure~\ref{fig:gbjmet} shows the $\met\;\;$ and dijet mass 
$M_{bj}$ distributions from data and predicted background. 
Other kinematic distributions, such as jet multiplicity, \et\ of photon, 
\et\ of $b$-tagged jet, {\it etc.}, have also been examined and no significant 
excess has been found. 
The observed number of events in data is 617, which is consistent 
with the expected number of background events, $637\pm 139$. 

\section{\boldmath SEARCH FOR ANOMALOUS PRODUCTION OF $\ell\g b\met\;\;$ 
  and Measurement of SM $t\bar{t}\g$ Production Cross-section
  \label{sec:ttg}} 
\begin{figure}
\includegraphics[width=0.28\textwidth, angle=90]{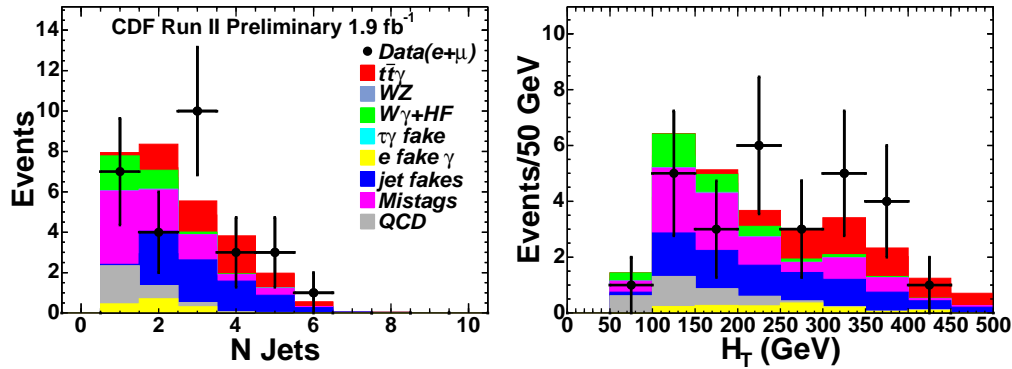}   
	\caption{\label{fig:ttg}
	CDF search for anomalous production of $\ell\g b\met\;$: 
	the jet multiplicity (left) and $H_T$ (right) distributions observed 
	(markers) and background prediction (filled histograms).  
	The contribution of SM $t\bar{t}\g$ increases as the jet multiplicity 
	and $H_T$ increase. }
\end{figure}

\unboldmath
Ref.~\cite{Kane:1996ny} predicts in the Minimal Supersymmetric 
Standard Model (MSSM) an exotic decay channel of the top quark, which may 
compete with $t\rightarrow Wb$, into a light stop and a light Higgsino-like 
neutralino. 
A $t\bar{t}$ pair may then decay via $t\bar{t}\rightarrow Wb\tilde t \tilde
 \chi_i \rightarrow \ell\bar{\nu}_{\ell}bc \kaione \kaione \g + X$. Instead of 
searching for this MSSM model only, the CDF collaboration has performed a 
model-independent search in the inclusive $\ell\g b\met\;$ final state using 
1.9~\fbarn\ of data, where $\ell$ is an electron or a muon. 
Since this signature is rare, the \et\ and $b$-tagging requirements are 
looser than those in Section~\ref{sec:gbjm}: a central electron or muon with 
$\pt>20~\gev$, a central photon with $\et>10~\gev$, at least one jet which is 
$b$-tagged by the loose SECVTX algorithm~\cite{SECVTX}, and $\met\; > 20~\gev$.
 Figure~\ref{fig:ttg} shows the jet multiplicity and $H_T$\footnote{The 
$H_T$ is defined as the scalar sum \pt\ of all identified objects, 
including $\met\;$, in an event.}
 distributions from the inclusive $\ell\g b\met\;$ final state. No significant 
excess in data is found: 28 observed and $27.9 {+3.6 \atop -3.5}$ 
expected. The background has a significant 
contribution from the SM $t\bar{t}\g$ production, especially in the lepton + 
jets channel. After requiring $H_T>200~\gev$ and two additional jets 
($\geq 3$ jets with $\geq$ 1 $b$-tag in total), the $t\bar{t}\g$ cross-section 
has been measured to be $0.15\pm0.08~\rm{pb}$, which is consistent with the 
next-to-leading-order (NLO) prediction, $0.080\pm0.012~\rm{pb}$~\cite{sigma_ttg}.

\section{ESTIMATION OF MIS-IDENTIFIED PHOTONS}
The number of events with mis-identified photons is estimated 
by the ``\textit{CES/CPR method}''~\cite{cescp2}, where we use 
cluster-shape variables from the central 
electromagnetic strip and wire chamber system (CES)~\cite{CES} and hit rates 
in the central preshower detector (CPR)~\cite{CPR}. The primary 
misidentified photons come from $\pi^0\rightarrow\gamma\gamma$ or 
$\eta^0\rightarrow\gamma\gamma$ decays where the hadrons are associated with 
jets. 
For photon candidates with $E_T<35$ GeV, the shape of the shower profile 
measured with CES can be used to discriminate between true single photon 
events and diphoton final states from decays of mesons. We construct a 
$\chi^2$ by comparing the measured shower profile with that from electron test 
beam data. A single photon has an average probability of $\sim80\%$ to satisfy 
a $\chi^2$ cut, while the background has an average probability of
$\sim30\%$, since the shower profile of the two near-by photons from a
meson decay is measurably wider on average. 
Above $35$ GeV, however, the two photons from $\pi^0$ decay coalesce and
the discrimination power of the shower profile measurement is lost.
In this $E_T$ range, we use hit rates in the CPR system to 
discriminate between single photons and diphotons from meson decays. A
single photon will convert and leave a hit in the preshower detector
with a probability of $\sim65\%$.  Backgrounds that decay into two
photons have a hit probability of $\sim85\%$ because the probability
that neither photon converts is lower than the probability that a
single photon does not convert. The difference of probabilities
between signal and background forms the basis of a statistical method
which assigns each event a weight for being a mis-identified photon. The weight
is a function of the energy of the photon candidate, the angle of
incidence, the number of primary interactions found in the event, the
shower profile $\chi^2$, and whether or not the photon candidate
leaves a hit in the preshower detector. The number of background photons,  
$N_\mathrm{bkg}$, is a sum of these weights:
\begin{equation}
N_\mathrm{bkg} = \sum_i W_i = \sum_i \frac{\epsilon^i_\mathrm{sig}-\epsilon^i}{\epsilon^i_\mathrm{sig}-\epsilon^i_\mathrm{bkg}},
\end{equation}
where $\epsilon^i$ is one or zero depending on if photon candidate $i$ satisfies the CES/CPR requirement or not\footnote{CES shower profile $\chi^2 < 4$ 
for $E_T < 35$~GeV and CPR ADC counts above threshold for $E_T \geq 35$~GeV.}. 
The $\epsilon^i_\mathrm{sig}$ and $\epsilon^i_\mathrm{bkg}$ are signal and 
background efficiencies given the kinematic 
information of photon candidate $i$ and are parameterized 
using subsidiary measurements. The ``CES/CPR'' method has been used 
in the measurement of inclusive photon and diphoton cross-sections 
and the search described in Section~\ref{sec:gbjm}. It may be 
applied in the measurement of $\gamma b\bar{b}$ cross-section and 
other searches in the future.

\section{CONCLUSION}
The CDF collaboration has performed signature-based searches in the 
$\gamma b \met\;$ + $X$ final state, where $X$ is an additional jet, 
an electron or a muon. 
We have not yet found significant excess in 1.9--2.0~\fbarn\ of data.  
We have measured the Standard Model $t\bar{t}\gamma$ cross-section 
to be $0.15\pm0.08~\rm{pb}$. 
The current measurement has a large statistical uncertainty and may be 
improved as more data data are being collected. Finally, 
we briefly describe the ``CES/CPR'' method which is used to estimate 
the contribution of mis-identified photons. This method may 
also be used in the future measurements of cross-sections and 
searches for new physics.


\begin{thebibliography}{99}
\bibitem{sigma_ttg} 
The leading-order (LO) production cross-section of $t\bar{t}\g$ has been 
estimated using the {\sc Madgraph} generator to be 0.076~pb. A $k$-factor to 
scale the LO to the NLO cross-section, $1.10\pm 0.015$, is used after a 
discussion with F.~Petriello and U.~Baur.
\end{thebibliography}
\end{document}